\magnification1200

\rightline{KCL-MTH-02-08}
\rightline{hep-th//0204207}

\vskip .5cm
\centerline
{\bf Massive IIA Supergravity as a Non-linear Realisation}
\vskip 1cm
\centerline{ Igor Schnakenburg and Peter West }
\vskip .5cm
\centerline{Department of Mathematics}
\centerline{King's College, London, UK}

\leftline{\sl Abstract}
\noindent 
A description of the bosonic sector of massive IIA supergravity as a
non-linear realisation is given.  An essential feature of this
construction is that the momentum
generators have non-trivial commutation relations  with the 
generators associated with the gauge fields. 

\vskip .5cm

\vfill

Igor Schnakenburg is financially supported by DAAD (D/00/09914).
\par
\vskip 1cm
email: schnake, pwest@mth.kcl.ac.uk

\parskip=18pt

\eject

%\pageno=1

\medskip
{\bf {0. Introduction }}
\medskip
Long ago Nahm [1] pointed out that supergravity theories could only
exist in eleven and less space-time dimensions and that the maximum
number of supersymmetries they could possess were contained in spinors
that had in total no more than 32 real  components.
The supergravity theories with 32 components are called maximal
supergravity theories. There is a unique such theory in eleven
dimensions which was constructed [2] as an  application
of the Noether method that was used to construct the first
supergravity theory,  the $N=1$, $D=4$ supergravity theory. However,
in ten dimensions there are two such maximal supergravity theories
called IIA and IIB. These two theories possess different
supersymmetries which when decomposed in terms of Majorana-Weyl
spinors are of opposite and the same chirality respectively. While the
construction of the IIA theory [3] was found by dimensional
reduction of the eleven dimensional supergravity theory, the
construction of the IIB theory [4,5,6] required new techniques. There is 
a modification of the IIA theory that preserves the number of
supersymmetries, but introduces a dimensionfull parameter [7]. This theory
is called  massive IIA supergravity and it possesses a cosmological
constant.   
\par
One of the most remarkable features of supergravity theories is that 
the scalars in the supergravity multiplets  always occur in a
coset structure [8]. While this  can be viewed as a consequence of
supersymmetry, the groups that occur in these cosets are rather
mysterious [9,10,11].  It has been conjectured [12] that the symmetries
found in these  cosets are symmetries of the associated  non-perturbative
string theory.  
\par
The coset construction for the description of the scalars was extended 
 to include the gauge fields for the maximal supergravity theories [13].
In this construction  all gauge fields and scalars were introduced along
with their duals. The advantage of this approach is that the equations of
motion can be reduced to first order equations in form of a
generalised self-duality condition. This method was subsequently
applied  to massive IIA [14]. However, in these papers the vector indices
on the gauge fields did not arise from the underlying  group, but were
introduced by hand. As such, it is difficult to see how this
construction could be extended to include the other degrees of
freedom of the theory, namely  the graviton and the fermions. 
\par
Recently, it was shown that the entire bosonic sectors, including
gravity, of the eleven dimensional supergravity theory and the ten
dimensional IIA  [15]  and IIB supergravity theories [16], 
could be formulated as a non-linear realisation. It was also 
conjectured, that when suitably formulated  eleven dimensional   
supergravity would be invariant under a Kac-Moody algebra [17]. Although,
this conjecture was not proved in [17] some evidence was given and the
Kac-Moody algebra was identified.   It had rank eleven and was denoted
$E_{11}$  [17]. A similar analysis  found that  $E_{11}$ was also the
Kac-Moody algebra that would underlie the ten dimensional IIA and IIB
theories [16,17]. As has been pointed out  by one of the authors (PCW) in
a number of seminars it could be that  $E_{11}$  is part of an even larger
underlying algebra that is a Borcherds algebra.  
\par
This is consistent with the general  belief  that
all these theories are part of a larger theory  which has been called M
theory [18]. It was conjectured [17] that 
$E_{11}$ over an appropriate field was a symmetry of M
theory. However, as seen from  this perspective M theory does not
necessarily live in eleven dimensions, but rather has its dimension
undetermined. A particular theory results from a choice of "vacuum" in M
theory and the dimension of the resulting theory is a consequence of the
Lorentz group that is contained in the subgroup of $E_{11}$ that is
preserved by the "vacuum" under consideration.   

In this paper we extend some of these considerations to the massive IIA
theory and  as a first step show that its entire bosonic sector
can be described as a non-linear realisation. 

%%%%%%%%%%%%%%%%%%%%%%%%%%%%%%%%%%%%%%%%%%%%%%%%%%%%
%
%    section {1. MASSIVE IIA SUGRA}
%
%%%%%%%%%%%%%%%%%%%%%%%%%%%%%%%%%%%%%%%%%%%%%%%%%%%%

\medskip
{\bf 1. Massive IIA Supergravity}
\medskip
The bosonic sector of massive IIA supergravity was originally [7]
described in terms of the same fields as in the ten-dimensional chiral
theory without the cosmological constant. These fields were  the
graviton, $h_a{}^b$, the scalar $A$ (dilaton) and the
$p$-forms $A_{a_1\cdots a_p}$ for $p= 1,2,3$.  However, the massive
theory also contains  a constant which  is related
to the cosmological constant. The bosonic part of the original
Lagrangian given by Romans in [7] was
$$
   e^{-1}L = R - {1\over 2} \partial_\mu A\partial^\mu A   
   -{1\over 12} e^{-1/2A} F^{\mu\nu\rho\sigma}
   F_{\mu\nu\rho\sigma} -{1\over 3}e^{A} G^{\mu\nu\rho}
   G_{\mu\nu\rho} - m^2 e^{-3/2A} B^{\mu\nu} B_{\mu\nu} 
$$
$$- {1\over 2}  m^2e^{-5/2A}
  +{e^{-1}\over 6\cdot 48} (\epsilon^{\mu\nu\rho\sigma\mu_1\cdots\mu_6}
   (16\partial_\mu A'_{\nu\rho\sigma} \partial_{\mu_1} 
   A'_{\mu_2\mu_3\mu_4} B_{\mu_5\mu_6}  
$$
$$+16m \partial_{\mu}
   A_{\nu\rho\sigma} B_{\mu_1\mu_2} B_{\mu_3\mu_4} B_{\mu5\mu_6}   
+{36\over5}m^2 B_{\mu\nu} B_{\rho\sigma} B_{\mu_1\mu_2}
   B_{\mu_3\mu_4} B_{\mu_5\mu_6}) ,
\eqno(1.1)
$$
where 
$$
   G_{\mu\nu\rho} = 3 \partial_{[\mu} B_{\nu\rho]}\, , \quad {\rm and} 
   \quad F_{\mu\nu\rho\sigma} = 4(\partial_{[\mu} A'_{\nu\rho\sigma]} +
   6m B_{[\mu\nu}B_{\rho\sigma]}). 
\eqno(1.2)
$$
By redefining the fields according to
$$
   B_{\mu\nu} = A_{\mu\nu} + {2\over m} \partial_{[\mu}A_{\nu]}\,
   \quad A'_{\mu\nu\rho} = A_{\mu\nu\rho} - 6 A_{[\mu}A_{\nu\rho]}
   -{6\over m} A_{[\mu} \partial_\mu A_{\rho]},
\eqno(1.3)
$$
as explained in [19], it can be rewritten in the form
$$
   e^{-1}L = R - {1 \over 2} \partial_\mu A \partial^\mu A -
   {1 \over 12} e^{1/2A} F_{\mu\nu\rho\sigma}F^{\mu\nu\rho\sigma}
   - {1 \over 3} e^{-A} G_{\mu\nu\rho}G^{\mu\nu\rho} -
   e^{3/2A} F_{\mu\nu} F^{\mu\nu} 
$$
$$ 
   - {1 \over 2} m^2 e^{5/2A}
   + {e^{-1} \over 18} \epsilon^{\mu_1\cdots\mu_{10}} (
   \partial_{\mu_1} 
   A_{\mu_2\mu_3\mu_4} \partial_{\mu_5} A_{\mu_6\mu_7\mu_8}
   A_{\mu_9\mu_{10}} +  m \partial_{\mu_1} A_{\mu_2\mu_3\mu_4}
   A_{\mu_5\mu_6} A_{\mu_7\mu_8}A_{\mu_9\mu_{10}}
$$
$$
   + {9 \over 20} m^2 A_{\mu_1\mu_2} A_{\mu_3\mu_4} A_{\mu_5\mu_6}
   A_{\mu_7\mu_8}A_{\mu_9\mu_{10}} + 18
   m\partial_{\mu_1} A_{\mu_2\cdots \mu_{10}} ) ,
\eqno(1.4)
$$ 
where the gauge invariant field strengths are now given by
$$
   F_{\mu\nu} =  2 (\partial_{[\mu} A_{\nu]} + {1\over 2}m
   A_{\mu\nu})\, ,\quad {\rm and}\quad  
   G_{\mu\nu\rho} = 3 \partial_{[\mu} A_{\nu\rho]},
\eqno(1.5)
$$
$$
   F_{\mu\nu\rho\sigma} = 4 (\partial_{[\mu}
   A_{\nu\rho\sigma]} + 6 A_{[\mu} \partial_{\nu} A_{\rho\sigma]} + {3
   \over 2} m A_{[\mu\nu} A_{\rho\sigma]} ).
\eqno(1.6)
$$
The terms containing negative powers of $m$ form a total divergence and
can be dropped. In this second formulation the one form gauge field, 
which had been absorbed in the two form gauge field to make it massive
in the former formulation, appears explicity and  one can take
$m\to 0$  to find the Lagrangian of massless IIA supergravity in
a straightforward way.

Following [19], in this formulation  we may  
 treat 
$m$ as a dynamical field. The field equation for the new nine form gauge
fields states that $\partial _\mu m=0$ i.e. $m$ is a constant. While the 
field equation for $m$ sets the field strength of the nine form gauge
plus  a combination of other
forms in the theory equal to the epsilon symbol.  
We note that unlike for the other
gauge fields these equations are first order. In [19] it was argued that
one can derive the field equations for the purely bosonic sector of
massive IIA supergravity from a Lagrangian which does not contain the
mass parameter
$m$ at all. To obtain this formulation the field equation for $m$ was
plugged back into the Lagrangian (1.4). The advantage of this approach, 
as was explained  in [19] is that this theory then naturally couples to
an eight brane.
\par
References [13] and [14] introduced duals of all the original gauge
fields. As a result, the field equations for the gauge potentials could 
be reduced to first order. In the case of massive IIA supergravity
[14] the dual of the nine form gauge field was taken to be a ``minus
one form'' whose field strength was then dual to the ten form field
strength of the nine form gauge field. We have noted in the last
paragraph that the field equation for the nine form gauge potential is
necessarily first order. We therefore do not introduce a dual of the nine
form potential. In this paper we will introduce dual gauge fields for all
the original gauge field, namely $A_{a_1\cdots a_q}$ for $q=5,6,7,8$, but
not for the nine  form gauge field. We will find that the momentum
operator plays an important role in place of the generator associated
with the minus one form. Although to see the full symmetry one will have 
to introduce ``duals of gravity'' we will not do this here. The complete
bosonic field content we require is thus given by:  
$$
   h_a{}^b, \, A,\,A_{c}, \, A_{c_1 c_2}, \, A_{c_1c_2
   c_3}, \, A_{c_1 \ldots  c_5},\, A_{c_1 \ldots c_6},\, A_{c_1\ldots
   c_7},\, A_{c_1 \ldots  c_8},\, A_{c_1 \ldots  c_9}.
\eqno(1.7)
$$
In a non-linear realisation these fields are considered as the
Goldstone bosons. We therefore introduce the corresponding generators
$$
   K^a{}_b,\, R, \, R^{c}, \, R^{c_1 c_2}, \, R^{c_1c_2
   c_3}, \,  R^{c_1\ldots  c_5},\, R^{c_1 \ldots c_6},\, R^{c_1\ldots
   c_7}, \, R^{c_1\ldots  c_8},\, R^{c_1 \ldots  c_9}.
\eqno(1.8)
$$
\par
The generators $K^a{}_b$ satisfy the commutation relations of
$GL(10,{\bf{R}})$ and the non-zero commutation relations between all
generators mentioned above are given by
$$
   [{K^a}_b ,  {K^c}_d] = \delta^c_b{K^a}_d -\delta^a_d{K^c}_b  ,
   \quad [{K^a}_b , P_c] = - \delta^a_cP_b, \quad
   [{K^a}_b  ,   R^{c_1\cdots c_p}] = \delta^{c_1}_bR^{ac_2\cdots
   c_p} + \cdots , 
$$
$$
   [R, R^{c_1\cdots c_p} ] = c_p R^{c_1\cdots c_p} \, , \quad
   [R^{c_1\cdots c_p} \, , \,  R^{c_1\cdots c_q}] =  c_{p,q}
   R^{c_1\cdots c_{p+q}},
\eqno(1.9)
$$ 
where $ + \cdots $ means the appropriate anti-symmetrisations. We 
also include the momentum generator $P_a$ in the symmetry algebra and
its non-trivial relations with the other generators are given by 
$$
   [P_a, R^{c_1\cdots c_p} ]= -m b_p (\delta_a^{c_1} R^{c_2\cdots c_p}
   + \cdots )\, , \quad [ P_a, R] = -mb_0 P_a.
\eqno(1.10) 
$$  
In what follows it will often be useful to denote $c_{-1}=mb_0$, since
in this way one can view the last commutator of equation (1.10) as an
extension of those of equation (1.9). 
\par
If the coefficients are taken to be
$$
   c_3 = - c_5 = -{1 \over 4},\quad c_2 = - c_6 = {1\over 2},\quad
   c_1 = - c_7 = -{3 \over 4},\quad c_9 = - c_{-1}= {5 \over 4},
$$
$$
   c_{1,2} = - c_{2,3} = -c_{3,3} = c_{1,5} = c_{2,5} = 2, \quad
   c_{3,5} = 1 ,\quad c_{2,6} = 2 ,\quad c_{1,7} = 3 ,
$$
$$
   c_{2,7}= - 4 ,\quad b_2 = - {1\over 2},\quad b_7 = -{1\over 2},
   \quad b_9 = {5\over 8}, 
\eqno(1.11)
$$
(all not mentioned coefficients are equal to zero) then we can 
verify that  all Jacobi identities are satisfied. For example, the
generators corresponding to the gauge fields (all $c$'s) fulfill the
condition 
$$
   c_{q,r} c_{p, q+r} = c_{p,q} c_{p+q,r} + c_{p,r} c_{q, p+r}
\eqno(1.12)
$$
where $q,p$ and $r$ indicate the rank  of the  generators. The 
Jacobi identities which involve $b_0$ obey the above relation
provided we take $mb_0=c_{-1}$ where appropriate. The Jacobi
identities that involve the $b_p, p\not= 0$ structure constants in
the  new commutators of equation  (1.10) obey the relation   
$$  
   b_{p+q} c_{p,q} = b_p c_{p-1,q} + b_q c_{p,q-1}.
\eqno(1.13)
$$ 
One such example is given by the Jacobi identity involving 
$R^{ab}$,  $R^{c_1\cdots c_7}$ and $P_d$ which is satisfied 
provided $ c_{2,7}b_9=c_{2,6}b_7 + b_2 c_{1,7}$.
\par
The relations of equations (1.9) are the same as the algebra denoted
$G_{IIA}$ [16] relevant to the IIA supergravity theory except that
they also include the rank nine generator. For the IIA theory the
commutators of equation (1.10) vanish. Since the above mentioned
commutation relations include those of IIA supergravity [16], we call
the modified group which is generated by the above generators
$G_{mIIA}$. 
\par
We can write a general group element of the corresponding  group as
$$
   g = exp (x^\mu P_\mu ) exp (h_a{}^b K^a{}_b ) g_A \equiv g_x g_h
   g_A, 
$$
where 
$$
   g_A = e^{(1/9!)A_{a_1 \cdots a_9}R^{a_1 \cdots a_9} }
      \, e^{(1/8!)A_{a_1 \cdots a_8}R^{a_1 \cdots a_8} }   
      \, e^{(1/7!)A_{a_1 \cdots a_7}R^{a_1 \cdots a_7} } 
 \times
$$
$$
   e^{(1/6!)A_{a_1 \cdots a_6}R^{a_1 \cdots a_6} }
   e^{(1/5!)A_{a_1 \cdots a_5}R^{a_1 \cdots a_5} } \, 
   e^{(1/3!)A_{a_1 a_2 a_3}R^{a_1 a_2 a_3} } \, 
   e^{(1/2!)A_{a_1 a_2}R^{a_1 a_2} } \, 
   e^{A_{a_1} R^{a_1}} \,
   e^{A R}.
\eqno(1.14)
$$
Of course, we could have chosen any other representation, but the
calculations turn out to be simplest in this particular exponential
representation.
\par
We now construct a non-linear realisation of the $G_{mIIA}$ algebra 
taking the local sub-algebra to be the Lorentz group. As such, we demand
that the theory is invariant under 
$$
   g \to g_0 g h^{-1},
\eqno(1.15)
$$
where $g_0$ is a rigid element from the whole group $G_{mIIA}$ and $h$
 is  a local Lorentz transformation.  
We
calculate the Maurer-Cartan form  
$$
   {\cal {V}} = g^{-1}dg -\omega
\eqno(1.16)
$$
in the presence of the Lorentz connection $\omega ={1\over 2}
dx^\mu\omega_{\mu b}{}^a J^b{}_a$, which transforms as
$$
   \omega \to h\omega h^{-1} + hdh^{-1}.
\eqno(1.17)
$$
As a result, ${\cal {V}}$ transforms as 
$$
   {\cal {V}} \to h {\cal {V}}  h^{-1}.
\eqno(1.18)
$$
We split the Cartan form in gravity and gauge field parts
according to
$$
   {\cal {V}} = (g_h^{-1}dg_h -\omega ) + (g_A^{-1}dg_A + g_A^{-1}
   (g_h^{-1}dg_h) g_A - g_h^{-1}dg_h),
\eqno(1.19)
$$
and set the result to be equal to
$$
   {\cal {V}} \equiv dx^\mu (\hat e_\mu{}^a P_a + {\Omega_{\mu
   a}}^b{K^a}_b) + dx^\mu ( \sum_{p=1,\ p\not= 4}^9{1\over p!}
   e^{-c_{p-1}A} \tilde D_\mu
   A_{a_1\cdots a_p}R^{a_1\cdots a_p}),
\eqno(1.20)
$$
where 
$$
   {\Omega_{ab}}^c \equiv {(\hat e^{-1})_a}^\mu \left(
   {(e^{-1}\partial_\mu e)_b}^c  - {\omega_{\mu b}}^c\right),
\eqno(1.21)
$$
and
$$
   \hat e_\mu{}^a= e^{-5/4A} (e^h)_\mu{}^a,\qquad 
   e_\mu{}^a= (e^h)_\mu{}^a .
\eqno(1.22)
$$
We see that the object  in front of the momentum generator gets
altered with respect to IIA (where it was $e_\mu{}^a$) due to the
non-vanishing commutator in equation (1.10). We also see that the
vielbeins in the $\Omega_{\mu b}{}^c$ equation (1.21) are the unhatted 
vielbeins. The additional factor of $e^{-5/4A}$ just multiplies
the usual expression of $\Omega_{ab}{}^c$ from the massless case
[16]. As we will see in chapter 2,  the physical vielbein therefore
remains unchanged. The objects $\tilde D_\mu A_{a_1\cdots a_p}$
defined in (1.20) will be explicitly stated below. 
\par
Massive IIA supergravity is the non-linear realisation of the group
that is the closure of the $G_{mIIA}$ algebra given above and the
conformal group in ten dimensions. We therefore take only those
combinations of the Cartan forms of $G_{mIIA}$ that can be rewritten
as Cartan forms of the conformal group (see section 2). Lorentz
covariant objects which are also covariant under the full non-linear
realisation of the closure of the conformal and the $G_{mIIA}$ algebra
are then for example the completely anti-symmetrised derivatives
$$
   \tilde F_{a_1\cdots a_p} = pe^{-5/4A} e^{-c_{p-1}A}\tilde   
   D_{[a_1}A_{a_2\cdots a_p]}, 
\eqno(1.23)
$$ 
where we have to use $\tilde D_a = \hat e_a{}^\mu\tilde D_\mu$ to
convert the curved index to a flat index group-covariantly. As the
physical vielbein is the unhatted one, we gain an additional factor
$e^{-5/4A}$ in front of every field strength in comparison with
the IIA case. A discussion of the closure with the conformal group is  
postponed to the second section. We now give the explicit form of the
field strengths. They are given for the scalar by:
$$
   \tilde F_a = e^{-5/4 A} D_a A,
\eqno(1.24)
$$
for the 1-form:
$$
   \tilde F_{a_1a_2} = 2e^{-5/4A} e^{(3/4) A}\tilde D_{[a_1}
   A_{a_2]}=2e^{-5/4A} e^{(3/4) A} 
   ( D_{[a_1} A_{a_2]} + {1\over 2} m A_{a_1a_2}), 
\eqno(1.25)
$$
for the 2-form:
$$
   \tilde F_{a_1a_2a_3} = 3e^{-5/4A} e^{-(1/2) A} D_{[a_1}
   A_{a_2a_3]} 
\eqno(1.26)
$$
for the 3-form:
$$
   \tilde F_{a_1\cdots a_4} = 4e^{-5/4A} e^{(1/4) A} (
   D_{[a_1} A_{a_2\cdots a_4]} + 6 A_{[a_1} D_{a_2} A_{a_3a_4]}
   + {3 \over 2} m A_{[a_1a_2} A_{a_3a_4]} ) 
\eqno(1.27)
$$
for the 5-form:
$$
   \tilde F_{a_1\cdots a_6} = 6 e^{-5/4A} e^{-(1/4)A}
   \left(  D_{[a_1} A_{a_2\cdots a_6]} + 20 ( D_{[a_1}
   A_{a_2\cdots a_4}  + {1 \over 2} m A_{[a_1a_2} A_{a_3a_4})
   A_{a_5a_6]}\right)  
\eqno(1.28)
$$
for the 6-form:
$$
   \tilde F_{a_1\cdots a_7} = 7 e^{-5/4A} e^{(1/2) A}\left(
   D_{[a_1} A_{a_2\cdots a_7]} + {1 \over 7} m
   A_{a_1a_2\cdots a_7} - 20
   A_{[a_1a_2a_3} D_{a_4} A_{a_5a_6a_7]} \right. +
$$
$$
   \quad+ \left. 12 A_{[a_1}( D_{[a_2} A_{a_3\cdots a_7]} +
   20 A_{a_2a_3}( D_{[a_4} A_{a_5a_6a_7]} + {1 \over 2}
   mA_{a_4a_5} A_{a_6a_7]})) \right)
\eqno(1.29)
$$
for the 7-form:
$$
   \tilde F_{a_1\cdots a_8} = 8 e^{-5/4A} e^{-(3/4)A}
   \left( D_{[a_1} A_{a_2\cdots a_8]} - 42 A_{[a_1a_2}(
   D_{a_3}  A_{a_4\cdots a_8]} + 10 D_{a_3} A_{a_4a_5a_6}
   A_{a_7a_8]} + \right. 
$$
$$
   + \left. {5 \over 2} m A_{a_3a_4}A_{a_5a_6}
   A_{a_7a_8]}) \right)
\eqno(1.30)
$$
for the 8-form:
$$
   \tilde F_{a_1\cdots a_9} = 9 e^{-5/4A}\left( D_{[a_1}
   A_{a_2\cdots a_9]} - 24 A_{[a_1}  D_{a_2} A_{a_3\cdots a_9]}
   +  56 A_{[a_1a_2} D_{a_3} A_{a_4\cdots a_9]} - \right.
$$
$$
   - 56 A_{[a_1a_2a_3} D_{a_4} A_{a_5\cdots a_9]}  + 1008
   A_{[a_1} A_{a_2a_3} D_{a_4} A_{a_5\cdots a_9]} + 8 m 
   A_{[a_1a_2} A_{a_3\cdots a_9]}  + 
$$
$$
   + {7! \over 2} m A_{[a_1} A_{a_2a_3}
   A_{a_4a_5} A_{a_6a_7} A_{a_8a_9]}  + \left. {8! \over 4}A_{[a_1}
   A_{a_2a_3} A_{a_4a_5} D_{a_6} 
   A_{a_7a_8a_9]} -  \right. 
$$
$$
   \left. - 1120 A_{[a_1a_2} A_{a_3a_4a_5} \tilde D_{a_6}
   A_{a_7a_8a_9]} - {5 \over 36} m A_{a_1a_2\cdots a_9} \right)
\eqno(1.31)
$$
and finally for the 9-form:
$$
   \tilde F_{a_1\cdots a_{10}} = 10 e^{-5/4A} e^{-(5/4) A}\left(
   D_{[a_1} A_{a_2\cdots a_{10}]} - 144 A_{[a_1a_2}
   D_{a_3} A_{a_4\cdots a_{10}]} + \right.
$$
$$
   + 3024  A_{[a_1a_2} A_{a_3a_4} D_{a_5} A_{a_6\cdots a_{10}]}
   +  \left. {9! \over 18} A_{[a_1a_2} A_{a_3a_4} A_{a_5a_6}
   D_{a_7} A_{a_8a_9a_{10}]} + \right.
$$
$$ 
   + \left. 3024 m A_{[a_1a_2}
   A_{a_3a_4} A_{a_5a_6} A_{a_7a_8}A_{a_9a_{10}]}  \right),
\eqno(1.32)
$$
where $D_a$ is the  covariant derivative 
$$
   D_a A_{a_1\ldots a_p}=e_a{}^\mu(\partial _\mu A_{a_1\ldots a_p} +
   (e^{-1}\partial_\mu e)_{a_1}{}^c A_{ca_2\ldots a_p}+ \ldots  )
\eqno(1.33)
$$
and $\ldots$ indicates the terms where $(e^{-1}\partial_\mu e)$ acts
on the other indices of the gauge field.
Also, we have written the exponential $e^{-5/4A}$ separately in front
of every field strengths to indicate that it is common to {\it all} of
them. 
\par
Using the Cartan forms which transform only under the local Lorentz group 
in a manner that their indices suggest we must write down a set of
invariant equations.  If we ask that they be first order in derivatives
they can only be  given by   
$$
  \tilde  F^{a_1\cdots a_p} = { 1 \over (10- p)!} \epsilon^{a_1\cdots 
  a_{10}} \tilde   F_{a_{p+1} \cdots a_{10}}, \quad\quad p = 1,2,3,4. 
\eqno(1.34)
$$
Here we see that the common exponential factor $e^{-5/4A}$ indeed
peels off each equation. The nine form gauge field does not possess a
dual field, however, its ten form field strength can be taken to be a
constant $m$ times the epsilon symbol 
$$
   m = e^{-5/4A}{1\over 10!} \epsilon^{a_1\cdots a_{10}}
   \tilde F_{a_1\cdots a_{10}}.
\eqno(1.35)
$$
All the above equations of motion and the Einstein equation are
equivalent to those one can derive from the Lagrangian formulation
given at the beginning of this chapter. We can see that the simple
field strengths of equations (1.5),(1.6) indeed match with those given
in our group approach of equations (1.25)-(1.27) and one can indeed
verify that the above equations of motion for the gauge sector (1.34) and
(1.35) are the same as those one can derive from the Lagrangian (1.4)
once we take $m$ to be a dynamical field.   
\par
To recover  the massless case, we simply switch
off the commutators with the momentum generator by setting $m=0$. 
However,  because  $c_{-1}=mb_0$ we also have $c_{-1} =0$. Then  using 
the Jacobi relation 
$$
   c_{-1} = - c_{9}
\eqno(1.36)
$$
we deduce $c_9=0$, and then as 
$$
   c_{2,7}c_9 = c_{2,7} ( c_2 + c_7),
\eqno(1.37)
$$
we also need $c_{2,7}= 0$. Thereby the nine form is made redundant
($\tilde F_{(10)} = dA_{(9)}$)  and we are
indeed left with the massless case.  

%%%%%%%%%%%%%%%%%%%%%%%%%%%%%%%%%%%%%%%%%%%%%%%%%%%%
%
%   section {2. Conformal group}
%
%%%%%%%%%%%%%%%%%%%%%%%%%%%%%%%%%%%%%%%%%%%%%%%%%%%%
\medskip
{\bf 2. Closure with the Conformal Group}
\medskip
The massive IIA supergravity theory is the non-linear realisation of 
the group that is the closure of the $G_{mIIA}$ group given above with 
the ten-dimensional conformal group. The closure of these two groups  
is an infinite dimensional group, but rather than working with this
group we can perform a simultaneous realisation of the $G_{mIIA}$  and
the conformal group. What this actually means is that we construct the 
equations of motion only from  combinations of the Cartan forms of the
$G_{mIIA}$ group, given above, that can be rewritten in terms of the
Cartan or other covariant forms of the conformal group. In doing
this one gains  invariance under both conformal group and
$G_{mIIA}$, and so necessarily we find invariance under the group
which is the closure of $G_{mIIA}$ and the conformal group. This is
discussed at length in reference [15], but here we briefly discuss the
novel features that arise in this procedure when applied to the
massive IIA theory.   
\par
The two groups only have one Goldstone boson in common namely the trace
of $h_a{}^b$ which is related to the conformal field $\sigma$. In fact we
have to identify these two fields via $e_\mu{}^a \equiv (e^h)_\mu{}^a
= (e^{\bar h + \delta\sigma})_\mu{}^a$ (as in [15]). All the other
fields that occur as Goldstone bosons in the $G_{mIIA}$ algebra are 
viewed as matter fields from the conformal group viewpoint. 
The {\bf conformal} covariant derivative of a field $B$
transforming under a representation $\Sigma$ of the Lorentz group is:  
$$
   \Delta_\mu B = e^{-\sigma}(\partial_\mu + \partial^\nu \sigma
   \Sigma_{\mu\nu})\, B.  
\eqno(2.1)
$$
In contrast, the {\bf $G_{mIIA}$} covariant derivative of a matter
field is given by
$$
   \tilde D_a B = (e^{-1})_a{}^\mu (\partial_\mu  + {1\over 2}
   \omega_{\mu b}{}^c \Sigma^b{}_c )\, B.
\eqno(2.2)
$$
multiplied by a suitable exponential of $A$. As the latter is a Lorentz
 scalar it plays no part for the discussion of the closure with the
conformal group given in this section. Solving (2.1) for
$\partial_\mu B$ and plugging the result into (2.2), we get
$$
   \tilde D_a B = (e^{-1})_a{}^\mu(e^\sigma \Delta_\mu B -
   \partial^\nu\sigma \Sigma_{\mu\nu} B + {1\over 2} \omega_{\mu
   b}{}^c \Sigma^b{}_c B ).
\eqno(2.3)
$$
If we demand that the whole $\sigma $ dependence be through the
conformal covariant derivative, then we learn from this equation that
$\omega_{\mu b}{}^c$ must be solved for  by a {\bf $G_{mIIA}$} 
 invariant condition on the Cartan forms in such a way as to   cancel
the  derivatives of
$\sigma$  on the right
hand side. This is carried out below and we find that 
 the usual expression of the spin connection in terms of the
vielbein (stated below in (2.8)) has precisely the right form to do the
job. 

We now illustrate the procedure of taking the closure of the conformal
and the $G_{mIIA}$ group by considering the vector field $A_a$ instead
of a general matter field $B$. The other gauge fields are treated in a
very similar way. The conformal covariant derivative of a vector (2.1)
is given by (see also [15])  
$$
   \Delta_\mu A_a =  e^{-\sigma}( \partial_\mu A_a + \eta_{\mu a}
   \partial^c\sigma A_c - \partial_a\sigma\delta_\mu^c A_c) .
\eqno(2.4)
$$
Using this equation we may  rewrite the $G_{mIIA}$ covariant
derivative of $A_a$ given in equation (1.25), respecting (1.33) as
$$
   \tilde  D_{a_1} A_{a_2} =  
\eqno(2.5)
$$
$$
   \bar e_{a_1}{}^\mu \left( \Delta_\mu
   A_{a_2} - e^{-\sigma} ( \eta_{\mu a_2}\partial^c\sigma A_c -
   \partial_{a_2} \sigma\delta_\mu^cA_c - \partial_\mu\sigma A_{a_2} -
   (\bar e^{-1} \partial_\mu \bar e)_{a_2}{}^c A_c - {1\over 2} m
   A_{\mu a_2} )\right) ,
$$
where the vielbein with the overbar stands for the traceless part  $
{e}_\mu{}^a = e^{\bar h_\mu{}^a+\delta_\mu^a\sigma} = (\bar e \cdot
e^{\sigma})_\mu{}^a$. We realise that only if we take the combination
$\tilde  D_{[a} A_{b]}$ does the  $\sigma$ dependence only appear 
 through the conformal covariant derivative alone, as then the 3
$\sigma$-dependent terms in the second bracket vanish since they are
symmetric in $\mu$ and $a_2$.  If we want 
to use expressions covariant under both the conformal group
 and $G_{mIIA}$, then we have to
demand that all $\sigma$ dependence be implicitly through the
conformal derivative alone. As such  we conclude that 
only the totally antisymmetrised object $2e^{3/4A}\hat
e_{[a|}{}^\mu\tilde  D_{\mu} A_{|b]} \equiv \tilde F_{ab}$ is covariant
under both groups. 
\par
We know that the closure of the $G_{mIIA}$ group and the conformal
group generates gauge transformations and general coordinate
transformations and so the above object should be covariant under
these transformations.  We observe that 
$$
   \tilde D_{[a}A_{b]} = 2\left( \partial_{[a}A_{b]} +
   (e^{-1}\partial_{[a} e)_{b]}{}^c A_c + mb_2 A_{ab}\right)
   =  2e^{}_a{}^\mu e^{}_b{}^\nu (\partial_{[\mu}
   A_{\nu]} + mb_2 A_{\mu\nu}),  
\eqno(2.6)
$$ 
making it clear that it is covariant under gauge and 
general coordinate transformations. The $G_{mIIA}$ covariant
derivatives of IIA supergravity only differ from those of massive IIA
supergravity by terms containing
$m$, and the nine form potential. However, as these new terms do not
contain derivatives the closure with the conformal group is not spoilt
by the presence of these terms.   
\par
We now turn to the gravity sector of the theory. Clearly, the
constraint 
$$
   \Omega_{a,[bc]} - \Omega_{b,(ac)} + \Omega_{c,(ab)} = 0.
\eqno(2.7)
$$
is $G_{mIIA}$ covariant, but one can show in much the same way 
as for the IIA and the eleven dimensional supergravity cases [15] that
it is also conformally covariant. Examining the definition of
$\Omega_{a,bc}$ in equation (1.21), we see that it involves an
undifferentiated factor of $\hat e_a{}^\mu = e^{-5/4A}
e_a{}^\mu$. The factor of $e^{-5/4A}$ can then be removed and we
find that it is exactly the same constraint as in the other cases
(eleven dimensional supergravity, IIA and IIB supergravity) and
therefore results in the usual expression for the spin connection in
terms  of the vielbein, namely 
$$
   \omega_{\mu bc} = {1\over 2} (e_b{}^\rho \partial_\mu e_{\rho c} -
   e_c{}^\rho\partial_\mu e_{\rho b}) -
   {1\over 2} ( e_b{}^\rho\partial_\rho e_{\mu c} -
   e_c{}^\rho\partial_\rho e_{\mu b}) -
   {1\over 2} ( e_b{}^\lambda e_c{}^\rho\partial_\lambda e_{\rho a} -
   e_c{}^\lambda e_b{}^\rho \partial_\lambda e_{\rho a}) e_\mu{}^a.
\eqno(2.8)
$$
We conclude that the physical vielbein of general relativity is just
$e_\mu{}^a$.  
\par
The upshot of this discussion is that the simultaneously covariant
objects that transform covariantly under  the group $G_{mIIA}$ {\it
and} the conformal group are the $\tilde F_{a_1\cdots a_p}$ for
$p=1,\ldots,10$ (except 5) defined in (1.24)-(1.32) and 
the Riemann tensor  
$$
   R_{\mu\nu b}{}^c \equiv \partial_\mu \omega_{\nu b}{}^c +
   \omega_{\mu b}{}^d \omega_{\nu d}{}^c - (\mu \leftrightarrow \nu) .
\eqno(2.9)
$$
The invariant field equations for all the fields except that of
gravity are  given  in equations (1.34) and (1.35), while that for
gravity must be of the form 
$$
   R_{\mu\nu b}{}^c e_c{}^\nu e_a{}^\mu  = {1\over 16} m^2
   e^{5/2A} \eta_{ab} + {1\over 2}
   \partial_a A\partial_b A 
   + e^{-A}( F^{(3)}_{a}{}^{cd}F^{(3)}_{bcd} - {1\over
   12} \eta_{ab}F^{(3)cdf}F^{(3)}_{cdf})
$$
$$ 
   + 2 m e^{3/2A} (F^{(2)}_a{}^{c}F^{(2)}_{bc}- {1\over 16}
   \eta_{ab}F^{(2)cd} F^{(2)}_{cd} ) + {1\over 3}
   e^{1/2A} (F^{(4)}_a{}^{cdf}F^{(4)}_{bcdf} -
   {3\over 32} \eta_{ab} F^{(4)cdfg}F^{(4)}_{cdfg}).
\eqno(2.10)
$$
This equation does not look $G_{mIIA}$ covariant as we seem to have
used the unhatted vielbeins only. However, the same factor of
$e^{-5/2A}$ turns up in every single term and can therefore again
be dropped. We note that, one cannot know the factor in front of each term
on the right-hand side. These factors can only be determined if we
additionally use information from supersymmetry or the Kac-Moody
algebra. We have just put in the correct values for those constants.  
\par

%%%%%%%%%%%%%%%%%%%%%%%%%%%%%%%%%%%%%%%%%%%%%%%%%%%%
%
%   section {3. Discussion }
%
%%%%%%%%%%%%%%%%%%%%%%%%%%%%%%%%%%%%%%%%%%%%%%%%%%%%
\medskip
{\bf 3. Discussion}
\medskip
We have shown that like all the other maximal supergravity theories
the entire bosonic sector of  massive IIA supergravity can also be 
described as a non-linear realisation. Apart from introducing dual
fields for all the gauge fields of the original formulation [7] of the  
massive IIA theory we also have included, following [19], a nine form
gauge which is associated with the introduction of the cosmological
constant. 
 The correct theory requires that the momentum generator has
non-trivial commutation relations with the generators associated with
the gauge fields as given in equation (1.10). This is natural as the
nine form is associated with the cosmological constant and so with
gravity.   
\par
This is in contrast to the work of reference [14] which takes a
different approach and does not include gravity, but  does introduce
a dual form for the nine form gauge field which was called a
``minus one form''. The properties of this minus one form are not very
explicitly spelt out. In effect we find in this paper that the
momentum generator plays the role of the generator associated with the
``minus one form'' of reference [14]. 
\par
It would be interesting to examine if   the non-linear realisation
could be extended, in ways explained in reference [17],  to be
invariant under a Kac-Moody, or Borcherds algebra, and to conjecture
what this algebra is. In the previous non-linear realisation of the
maximal supergravities the  momentum generator has not played a
central part in the Kac-Moody algebra that has been identified. 
 However, the non-trivial relations of equations (1.10)  imply that
this generator must occur in a non-trivial way in  the corresponding
algebra. Progress in this direction may also shed light on the place
that the massive IIA theory has in M theory.   

%%%%%%%%%%%%%%%%%%%%%%%%%%%%%%%%%%%%%%%%%%%%%%%%%%%%
%
%   section { Acknowledgements }
%
%%%%%%%%%%%%%%%%%%%%%%%%%%%%%%%%%%%%%%%%%%%%%%%%%%%%
\medskip
{\bf Acknowledgements:}
\medskip
IS would like to thank Andr\'e Miemiec, who has given support when 
calculating the equations of motion for various SUGRAs. IS is also
supported by DAAD (D/00/09914).

\medskip
{\bf {References}}
\medskip
\parskip 0pt
\item{[1]} W. ~Nahm, {\it "Supersymmetries and their representations"},
     Nucl. Phys. {\bf B135} (1978), p.149

\item{[2]} E.~Cremmer, B.~Julia, and J.~Scherk, {\it ``Supergravity theory
in
11 dimensions''},  Phys. Lett. {\bf B76} (1978) 409--412.

\item{[3]}
I.~C.~G. Campbell and P.~C. West, {\it ``N=2 d = 10 nonchiral supergravity
and
its spontaneous compactification''},  Nucl. Phys. {\bf B243} (1984)
112.;
  M. Huq and M. Namazie,
{\it ``Kaluza--Klein supergravity in ten dimensions''},
Class.\ Q.\ Grav.\ {\bf 2} (1985).;
  F. Giani and M. Pernici,
{\it ``$N=2$ supergravity in ten dimensions''},
Phys.\ Rev.\ {\bf D30} (1984) 325. 

\item{[4]} J.~H. Schwarz and P.~C. West, ``Symmetries and transformations
of chiral {N}=2,{D} = 10 supergravity,''  Phys. Lett. {\bf B126}
(1983) 301.

\item{[5]} J.~H. Schwarz, ``Covariant field equations of chiral {N}=2 {D}
= 10 supergravity,'' Nucl. Phys. {\bf B226} (1983) 269.

\item{[6]} P.~S. Howe and P.~C. West, ``The complete {N}=2, d = 10
supergravity,''  Nucl. Phys. {\bf B238} (1984) 181.

\item{[7]} L.~J. Romans, {\it ``Massive N=2a Supergravity in Ten
Dimensions,''}  Phys.\ Lett\  {\bf B169} (1986) 374.

\item{[8]} S.\ Ferrara, J.\ Scherk and B.\ Zumino, 
``Algebraic Properties of Extended Supersymmetry'',
Nucl.\ Phys.\ {\bf B121} (1977) 393;
E.\ Cremmer, J.\ Scherk and S.\ Ferrara, {\it ``SU(4) Invariant
Supergravity Theory''}, Phys.\ Lett.\ {\bf B74} (1978) 61.

\item {[9]} E. Cremmer and B. Julia,
{\it ``The $N=8$ supergravity theory. I. The Lagrangian''},
Phys.\ Lett.\ {\bf B80} (1978) 48
\item{[10]} B.\ Julia, {\it ``Group Disintegrations''},
in {\it Superspace \&
Supergravity}, p.\ 331,  eds.\ S.W.\ Hawking  and M.\ Ro\v{c}ek,
Cambridge University Press (1981).

\item{[11]}  H.  Nicolai, Phys. Lett. {\bf B187} (1987) 316.

\item{[12]} C.~M. Hull, P.~K. Townsend, {\it ``Unity of Superstring
  Dualities,''} Nucl. Phys. {\bf B438} (1995), 109, {\tt hep-th/9410167} .

\item {[13]} E.~Cremmer, B.~Julia, H.~L{\"u}, and C.~N. Pope,
  {\it ``Dualisation of dualities. I \& II: Twisted self-duality of
    doubled fields and superdualities, ''}  Nucl. Phys. {\bf B535}
  (1998) 242, {\tt hep-th/9806106}

\item{[14]} I.~V. Lavrinenko, H. L\"u, C.~N. Pope, K.~ Stelle, {\it
    ``Superdualities, Brane Tensions and Massive IIA/IIB Duality,''}
  Nucl.Phys. {\bf B555} (1999) 201, {\tt hep-th/9903057}

\item{[15]} P.~C. West, {\it ``Hidden superconformal symmetry in {M}
    theory ''},  JHEP {\bf 08} (2000) 007, {\tt hep-th/0005270}

\item{[16]} I. Schnakenburg, P. West, {\it ``Kac-Moody
Symmetries of IIB Supergravity,''}, Phys. Lett. {\bf B517} (2001)
421, {\tt hep-th/0107181}

\item{[17]} P. West, {\it ``E(11) and M theory},  
Classical and Quantum Gravity, {\bf 18} (2001) 4443, {\tt
hep-th/0104081}

\item{[18]} E. Witten, {\it String theory dynamics in various
dimensions}, Nucl. Phys. {\bf B443} (1995) 85, {\tt hep-th 9503124}

\item{[19]} E. Bergshoeff, M. ~de Roo, M.~B. Green, G. Papadopoulos,
  P.~K. Townsend {\it ``Duality of Type II 7-branes and 8-branes,''}
Nucl. Phys. {\bf B470} (1996) 113, {\tt 9601150}

%%%%%%%%%%%%%%%%%%%%%%%%%%%%%%%%%%%%%%%%%%%%%
\end